# *NFVactor*: A Resilient NFV System using the Distributed Actor Model

Jingpu Duan, Xiaodong Yi, Shixiong Zhao, Chuan Wu, Heming Cui, Franck Le

**Abstract**

Resilience functionality, including failure resilience and flow migration, is of pivotal importance in practical network function virtualization (NFV) systems. However, existing failure recovery procedures incur high packet processing delay due to heavyweight process-checkpointing, while flow migration has poor performance due to centralized control. This paper proposes *NFVactor*, a novel NFV system that aims to provide lightweight failure resilience and high-performance flow migration. *NFVactor* enables these by using actor model to provide a per-flow execution environment, so that each flow can replicate and migrate itself with improved parallelism, while the efficiency of the actor model is guaranteed by a carefully designed runtime system. Moreover, *NFVactor* achieves transparent resilience: once a new NF is implemented for *NFVactor*, the NF automatically acquires resilience support. Our evaluation result shows that *NFVactor* achieves 10Gbps packet processing, flow migration completion time that is 144 times faster than existing system, and packet processing delay stablized at around 20 microseconds during replication.

## I. Introduction

Network function virtualization (NFV) advocates moving *network functions* (NFs) out of dedicated hardware middleboxes and running them as virtualized applications on commodity servers [1]. To enable effective large-scale deployment of virtual NFs, a number of NFV management systems have been proposed in recent years [2], [3], [4], [5], [6], [7], implementing a broad range

J. Duan, X. Yi, S. Zhao, C. Wu and H. Cui are with the Department of Computer Science, The University of Hong Kong, email: (jpduan, xdyi, sxzhao, cwu, heming@cs.hku.hk).

F. Le is with the IBM T. J. Watson Research Center, email: (fle@us.ibm.com).

of management functionalities. Among these functionalities, resilience guarantee, supported by flow migration and failure recovery mechanisms, is of particular importance in practical NFV systems.

*Resilience to failures [8], [9] is crucial for stateful NFs.* Many NFs maintain important per-flow states [10]: IDSs such as Bro [11] store and update protocol-related states for each flow for issuing alerts for potential attacks; firewalls [12] parse TCP SYN/ACK/FIN packets and maintain TCP connection related states for each flow; load balancers [13] may retain mapping between a flow identifier and the server address, for modifying destination address of packets in the flow. It is critical to ensure correct recovery of flow states in case of NF failures, such that the connections handled by the failed NFs do not have to be reset – a simple approach strongly rejected by middlebox vendors [8].

*Efficient flow migration [14], [15], [16] is important for long-lived flows in case of dynamic system scaling.* Existing NFV systems [2], [6] mostly assume dispatching new flows to newly created NF instances when existing instances are overloaded, or waiting for remaining flows to complete before shutting down a mostly idle instance, which are efficient for short flows. Long flows are common on the Internet: a web browser uses one TCP connection to exchange many requests and responses with a web server [17]; video-streaming [18] and file-downloading [19] systems maintain long-lived TCP connections for fetching large volumes of data from CDN servers. When NF instances handling such flows are overloaded or under-loaded, migrating flows to other available NF instances enables timely hotspot resolution or system cost minimization [15].

Even though failure resilience and efficient flow migration are important for NFV systems, enabling light-weight failure resilience and high-performance flow migration within existing NF software architecture has been a challenging task.

Failure resilience in the existing systems [8], [9] is typically implemented through check-pointing: each NF process is regularly checkpointed, and if it fails, the system replays important log traces collected since the latest checkpoint to recover the failed NF. Compared to the normal packet processing delay of an NF that lies within tens of microseconds, the process of checkpointing is heavyweight and can cause extra delay up to thousands of microseconds [8], [9].

Flow migration in existing systems [14], [15] is typically governed by a centralized controller. It fully monitors the migration process of each flow by installing SDN rule to update the route of the flow and exchanging messages with the NFs over inefficient kernel networking stack [20]. However, a practical NFV system needs to manage tens of running NFs and handle tens of thousands of concurrent flows. To migrate these flows, the controller needs to sequentially execute the migration process of each flow, install a large number of SDN rules and exchange many migration protocol messages through inefficient communication channel, which may prolong the flow migration completion time and inhibit flow migration from serving as a practical NFV management task.

Besides, enabling flow migration with existing NF software is not trivial: OpenNF [15] reports that thousands lines of patch code must be added to existing NF software [11], [21] in order to extract and serialize flow states, communicate with the controller and control flow migration. This approach mixes the logic for controlling flow migration together with the core NF logic. To maintain and upgrade such an NF, the developer must well understand both the core NF logic and the complicated flow migration process, which adds additional burden on the developer.

In this paper, we present *NFVactor*, a new software framework for building NFV systems with high-performance flow migration and lightweight failure resilience. Unlike previous systems [8], [9], [14], [15] which augment existing NF software with resilience support, *NFVactor* explores new research opportunities brought by a holistic approach: *NFVactor* provides a general framework with built-in resilience support by exploiting the distributed actor model [22], and exposes several easy-to-use APIs for implementing NFs. Internally, *NFVactor* delegates the processing of each individual flow to an unique flow actor. The flow actors run in high-performance runtime systems, handle flow processing and ensure their own resilience in a largely decentralized fashion. *NFVactor* brings three major benefits.

▹ *Transparent resilience. NFVactor* ensures that once the NFs are correctly implemented with the provided APIs, failure resilience of the NFs is immediately achieved. *NFVactor* decouples resilience logic from core NF logic by incorporating resilience operations within the framework and only exposing APIs for building NFs. Using the APIs, programmers are fully liberated from reasoning about details of resilience operations, but only focus on implementing the processing logic of NFs and handling simple interaction for synchronizing shared states of NFs during resilience operations. The exposed APIs also ensure a clean separation between the core processing logic and important NF states, facilitating resilience operations.

▹ *Lightweight failure resilience.* With the actor abstraction and cleanly separated NF states, *NFVactor* is able to replicate each flow independently without checkpointing the entire NF process. Each flow actor can replicate itself by constantly saving its per-flow state to another actor that serves as a replica. This lightweight resilience operation guarantees good throughput, short recovery time and a small packet processing delay.

▹ *High-performance flow migration.* The use of the actor model enables *NFVactor* to adopt a largely decentralized flow migration process: each flow actor can migrate itself by exchanging messages with other flow actors, while a centralized controller only initiates flow migration by instructing a runtime system about the amount of the flow actors that should be migrated. As a result, *NFVactor* is able to concurrently migrate a large number of flows among multiple pairs of runtime systems. *NFVactor* also replaces SDN switch with a lightweight virtual switch for flow redirection, simplifying flow redirection from updating SDN rule into modifying an runtime identifier number. The increased parallelism and simplified flow redirection jointly enhance the performance of flow migration.

Our major technical challenge is to build an actor runtime system to satisfy the stringent performance requirement of NFV application. Even the fastest actor runtime systems [23] may fail to deliver satisfactory packet processing performance due to their actor scheduling strategies and the use of kernel networking stack. To address this challenge, we carefully craft a high-performance actor runtime system by combining the performance benefits of (i) a module graph scheduler to effectively schedule multiple flow actors, (ii) a DPDK-based [24] fast packet I/O framework [25] to accelerate network packet processing and (iii) an efficient user-space message passing channel which completely bypasses the kernel network stack and improves the performance of both failure resilience and flow migration.

We implement *NFVactor* and build several useful NFs using the exposed APIs. Our testbed experiments show that *NFVactor* achieves 10Gbps line-rate processing for 64-byte packets, concurrent migration of 600K flows using around 700 milliseconds, and recovery of a single runtime within 70 milliseconds in case of failure. Compared with OpenNF [15], flow migration completion time in *NFVactor* can be 144 times faster. Compare with FTMB [8] for replication performance, *NFVactor* achieves similar packet processing throughput and recovery time, but with packet processing latency stabilized at around 20 microseconds. The source code of *NFVactor* is available at [26].

## II. Motivations for Using the Actor Model

Systems that enable failure resilience [8], [9] and flow migration [14], [15] typically achieve a low level of parallelism: a centralized controller governs the migration of all the flows among multiple NF instances, while an entire NF process has to be checkpointed for replication. If we can improve the parallelism by providing an efficient per-flow execution environment, then each flow can migrate and replicate itself without full-process checkpointing and centralized migration control, leading to improved resilience performance. Such a per-flow execution environment can be modelled by actors.

The actor programming model [22], [27], [28], [29] has a long history of being used to construct massive, distributed systems [22], [28], [30], [31]. Each actor is a lightweight and independent execution unit. In the simplest form, an actor contains a global unique address, a message queue (mailbox) for accepting incoming messages, several message handlers and an internal actor state (*e.g.*, statistic counter, number of outgoing requests). An actor can send messages to other actors by referring to their addresses, process incoming messages using message handlers, update its internal state, and create new actors. Multiple actors run asynchronously as if they were running in their own threads, simplifying programmability of distributed protocols and eliminating potential race conditions that may cause system crash. Actors typically run on a powerful runtime system [29], which can schedule millions of lightweight actors simultaneously.

The actor model is a natural fit to provide a per-flow execution environment for resilient NFV system. We can create one actor as the basic execution environment for a flow and equip the actor with necessary message handlers for service chain processing, flow state replication and migration. Then each actor can process network packets and handle its own resilience by creating new actors and exchanging messages with other actors.

There are several popular actor frameworks [28], [27], [32], [29], but none of these frameworks are optimized for building NFV systems. In our initial implementation, we built *NFVactor* on top of libcaf [29], one of the fastest actor system [23]. But the overall performance turned out to be less than satisfactory, due to its actor scheduling strategy and the use of kernel networking stack. This inspires us to customize a high-performance actor runtime system (Sec. VI) for *NFVactor*. When being compared with one of the fastest actor libraries [23] in section VII-A, the packet processing throughput of our customized runtime increases by over 100% due to the improved actor scheduling strategy, while the speed for transmitting small actor messages increases up to

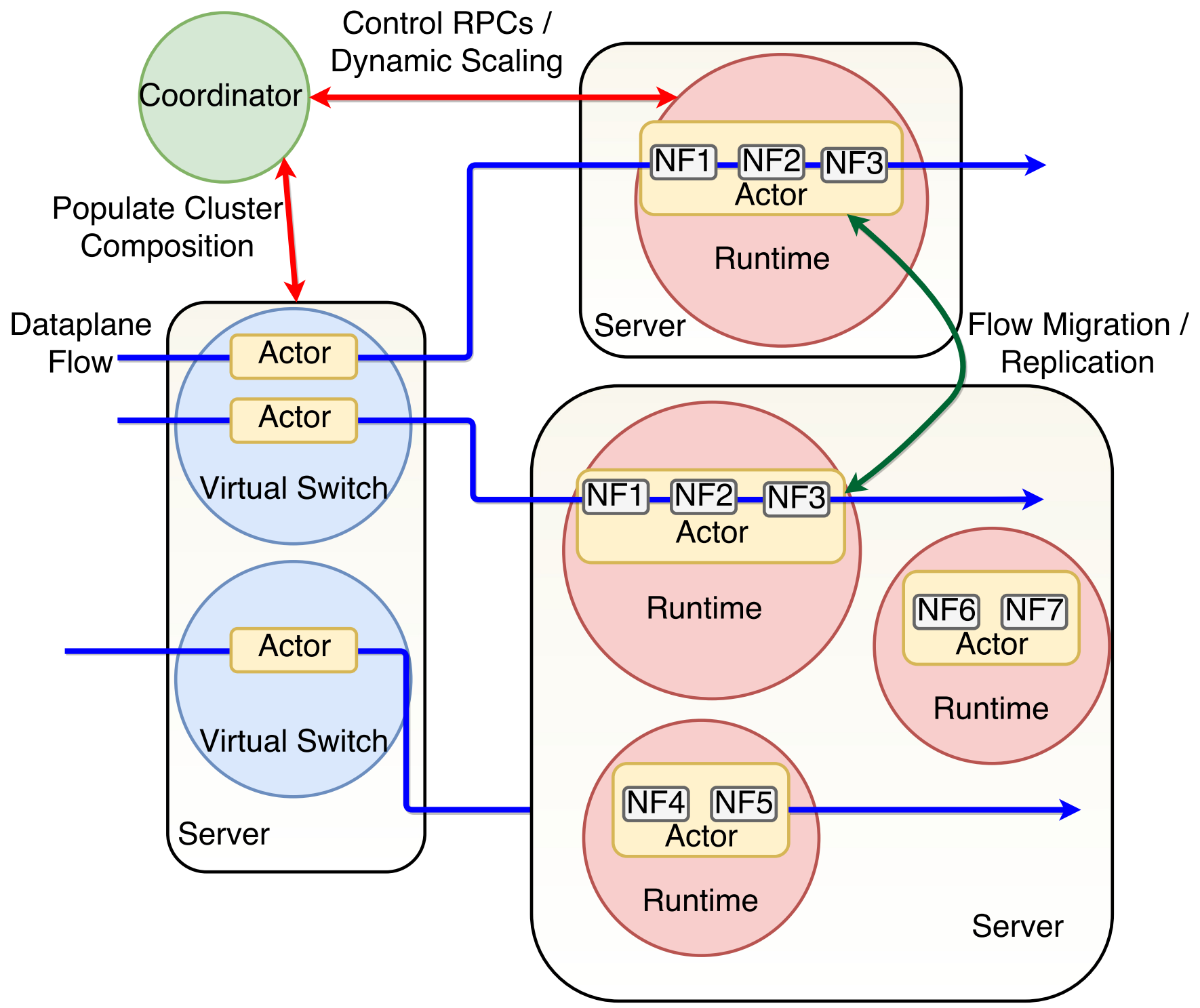

**Fig. 1:** An overview of the basic components of *NFVactor*. Three clusters are shown in this figure: a cluster for provisioning service chain ‘NF1→NF2→NF3’, a cluster for service chain ‘NF4→NF5’ and a cluster for service chain ‘NF6→NF7’

over 500% due to the removal of the kernel networking stack.

## III. The NFVactor Framework

### A. Overview

At the highest level, *NFVactor* has a layered structure as shown in Fig. 1. There are three key elements in *NFVactor*: (i) runtime systems (referred to as *runtime* for short) that enable flow processing using actors; (ii) virtual switches for distributing flows to runtime systems and sending flows to final destinations; and (iii) a lightweight coordinator for basic system management.

A runtime (Sec. III-B) is the execution environment of flow actors, running on a Docker container [33] for quick launching and rebooting, and is assigned a globally unique ID upon creation. A virtual switch is a special runtime (Sec. III-C) and serves as the gateway to runtimes. Runtimes and virtual switches are partitioned into several virtual clusters. In a cluster, the runtimes are initialized with the same service chain (Sec. III-B3) and the virtual switches dispatch flows to the runtimes within the same cluster. The partitioning of virtual clusters enables *NFVactor* to simultaneously provision multiple service chains.

Each virtual switch is configured with an entry IP address. The coordinator sets up proper switching rules to direct dataplane flows to virtual switches, which further dispatch them to runtimes within the same cluster. A runtime creates a dedicated flow actor to process each flow and forward flow packet to its final destination. The coordinator also manages dynamic scaling and failure recovery of *NFVactor* by interacting with runtimes and virtual switches through a series of control RPCs (Sec. III-D).

Dataplane flows can be migrated and replicated from one runtime to another runtime within the same cluster in a distributed fashion without persistent involvement of the coordinator. The details of flow migration and replication are further introduced in Sec. V.

### B. Runtime

*NFVactor* employs a carefully designed, uniform runtime system to run flow actors. Within a runtime, we adopt a *one-actor-one-flow* design principle: a dedicated flow actor is created to handle each flow received by the runtime. Packet processing by NFs and resilience operations are all implemented as reactive message handlers of the flow actor. The runtime timely schedules each flow actor to react to the various events, so that each flow actor can process flow packets and manage its own resilience in a largely decentralized fashion. Our one-actor-one-flow principle improves the parallelism of resilience operations and serves as the basis for high-performance resilience support.

*1) Internal Structure:* Fig. 2 shows the internal structure of a runtime. Each runtime is configured with one worker thread and one RPC thread. The worker thread actively polls the three ports shown in Fig. 2 and works in a run-to-completion mode. It is pinned to a dedicated CPU core to minimize the performance impact caused by thread scheduling. The RPC thread responds to RPC requests sent from the coordinator, for basic system management operations (Sec. III-D). The three ports are high-speed virtual NICs (ZeroCopyVPort in BESS [25]) and they are connected to a virtual L2 switch (L2Forward module of BESS) inside a physical server. The worker thread can bypass the kernel and directly fetch network packets from these ports.

*2) Work Flow:* The runtime has three basic work flows:

**Process Dataplane Flows:** The worker thread constantly polls dataplane flow packets from the input port. For each packet, the worker thread uses the 5-tuple of the packet (*i.e.*, source

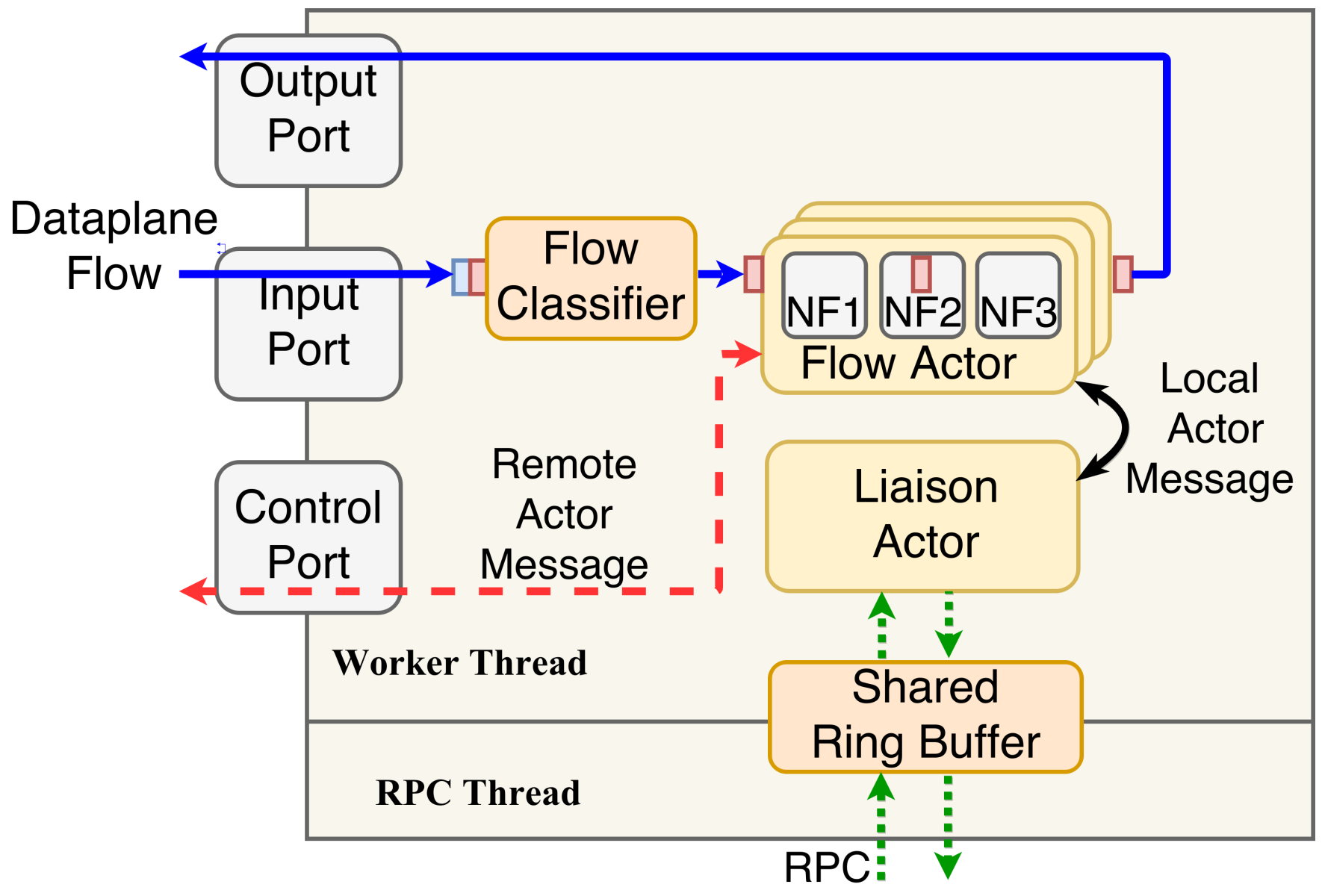

**Fig. 2:** The internal structure of a runtime.

IP address, destination IP address, transport-layer protocol, source port and destination port) to retrieve the corresponding flow actor and sends the packet to the flow actor for processing. Running in its own logical thread, the flow actor processes the packet along the configured service chain and then sends the processed packet out from the output port.

**Process Remote Actor Messages:** During distributed flow migration and replication, *remote actor messages* are exchanged among actors running on different runtimes. The runtime is equipped with a reliable message passing channel (Sec. VI) to reliably send and receive remote actor messages over the control port. The received remote actor messages are handed over to the destination actors for processing. The sent remote actor messages are reliably delivered to their receivers.

**Process Control RPCs:** The RPC thread forwards received RPC requests to a *liaison actor* in the worker thread through the shared ring buffer. The liaison actor coordinates with flow actors via *local actor messages*, to handle RPC requests sent from the coordinator.

*3) Service Chain:* Each runtime is configured with a sequential service chain (*e.g.*, firewall→ NAT→load-balancer) and initializes all the NFs along the service chain upon booting. When the flow actor processes packets, it calls the $process_pkt(input_pkt, fs, ss)$ API (Sec. IV) of each NF according to the service chain structure to implement the service chain processing logic.

### C. Virtual Switch

A virtual switch is a special runtime where the actors do not run service chains but only a flow forwarding function. An actor in a virtual switch is referred to as a *virtual switch actor*. The virtual switch serves as a load-balancing gateway when forwarding flows and a lightweight flow redirector when executing resilience operations.

For flow forwarding, a virtual switch learns runtimes that it can dispatch flows to through RPC requests sent from the coordinator. When a new flow arrives, a virtual switch actor selects a runtime with the smallest workload as the destination and saves its ID. For each flow packet, the virtual switch actor replaces the destination MAC address with the MAC address of the input port of the destination runtime and forwards the packet.

During flow migration and replication, each virtual switch actor can independently update the flow route by simply modifying the ID of the destination runtime. Compared with installing flow rules on an SDN switch [14], [15], the route update process is lightweight and improves flow migration performance for *NFVactor*.

### D. Coordinator

The coordinator in *NFVactor* is responsible for basic cluster management routines. As compared to centralized controllers in the existing NFV systems [15], [14], the coordinator only uses light-weight RPC calls to initiate the flow migration and replication process.

The coordinator communicates with runtimes via a number of control RPCs summarized in Tbl. I. It uses $poll_workload()$ to acquire the current workload on a runtime. It updates cluster composition (including MAC addresses of input/output/control ports, workload status and IDs of all runtimes and virtual switches in the cluster) to all the runtimes and virtual switches in a cluster using $notify_cluster_cfg(cfg)$.

To deploy a cluster, the system operator first specifies the composition of a service chain to the coordinator. The coordinator then creates a new cluster with one runtime and one virtual switch, configures the runtime with the specified service chain and installs proper switching rules to forward matching input flows to the virtual switch. The cluster is then dynamically scaled and recovered under the control of the coordinator.

The last three RPCs shown in Tbl. I are used to initiate flow migration and replication. After issuing these three calls, migration and replication are automatically executed among runtimes without further involving the coordinator.

**TABLE I:** Control RPCs Exposed at Each Runtime

| **Control RPC** | **Functionality** |
|---|---|
| poll_workload() | Poll the load information from a runtime. |
| notify_cluster_cfg(cfg) | Notify a runtime/virtual switch the current cluster composition. |
| set_migration_target(runtime_id, migration_number) | Initiate flow migration. It tells the runtime to migrate migration_num of flows to the runtime with runtime_id. |
| set_replicas(runtime_id_list) | Set the runtimes with IDs in runtime_id_list as the replica. |
| recover(runtime_id) | Recover all the flows replicated from runtime with runtime_id. |

**Dynamic Scaling.** The coordinator performs dynamic scaling of the runtimes and virtual switches by exploiting the distributed flow migration mechanism. The coordinator periodically polls the workload statistics from all the runtimes, containing the number of dropped packets on the input port, the current packet processing throughput and the number of active flows. In the current *NFVactor* prototype, the runtime is identified as overloaded if the number of dropped packets exceeds a fixed threshold (100 as in our experiments). This is an effective overload indicator for *NFVactor*: an overloaded runtime can not timely poll all the packets from its input port, therefore increasing the number of dropped packets significantly, while the CPU usage is rendered ineffective as the worker thread is a busy-polling thread and uses 100% of the CPU all the time.

If there is an overloaded runtime in a cluster, the coordinator launches a new runtime in the same cluster and keeps migrating a configurable number of flows from overloaded runtime (500 as in our experiments) to the new runtime, until half of the workload on the overloaded runtime is migrated away. If the new runtime becomes overloaded, more runtimes are added.We add new runtimes instead of moving flows across existing runtimes, since the load on existing runtimes

**TABLE II:** APIs for implementing NFs in *NFVactor*

| API | Usage |
|---|---|
| nf.allocate_shared_state() | Allocate a singleton object containing the shared states. |
| nf.allocate_new_fs() | Create and initialize a new flow state object. |
| nf.deallocate_fs(fs) | Deallocate the flow state object upon expiration of the flow actor. |
| ⋆ nf.process_pkt(input_pkt, fs, ss) | Process the input packet using the current flow states of the flow and the shared states of the NF. |
| nf.flow_expires(fs, ss) | Update the shared states according to final flow states upon expiration of the flow actor. |
| nf. flow_migrate_out(fs, ss)<br>nf. flow_migrate_in(fs, ss)<br>nf. flow_recover(fs, ss) | Update the shared states using the flow states during flow migration and replication. |

is largely balanced, due to the load-balancing functionality of virtual switches.

If runtimes in a cluster become largely idle, the coordinator carries out scale-in: it selects a runtime with the smallest throughput, migrates all its flows to the other runtimes, and shuts the runtime down when all the flows have been successfully moved out.

## IV. NF APIs

To create an NF with full resilience support, the programmer should properly implement the APIs listed in Tbl. II and ensure that the implemented APIs correctly satisfy the usage description in Tbl. II. We follow two principles when designing these APIs.

*First*, StatelessNF [34] and Split/Merge [14] demonstrate that it is possible to build practical NFs by processing each individual flow with its per-flow state and shared state. Inspired by this principle, *NFVactor* employs a core API $process_pkt(input_pkt, fs, ss)$ to accomplish the core NF processing logic. It is called by each flow actor when processing the input packet, taking per-flow state and shared state as additional arguments. Several supporting APIs are also provided to manage important NF states. This design ensures a clean separation between useful NF states and the core processing logic of an NF, so that the flow actor always has direct and efficient access to the latest flow states to ease flow migration and replication.

*Second*, to properly handle shared state, we treat shared state accessing by an NF as allocating resource from a shared resource pool. For instance, when a NAT processes a flow, accessing shared state usually means allocating an address from a shared address pool. Therefore, when the flow expires, the resource that the flow acquired should be properly released back to the shared resource pool. With the NAT example, this means that the allocated address should be put back into the address pool when the flow expires. However, when the flow is migrated or recovered on another NF instance, without proper synchronization, the resource obtained by the flow may not be correctly released back to the shared resource pool. To resolve this issue in *NFVactor*, the programmer should properly store the allocated resource in the per-flow state. They also need to implement the last four APIs shown in Tbl. II for properly releasing the acquired resource so that the shared state is correctly synchronized. Our runtime guarantees that the three APIs are timely invoked during flow migration and replication (Sec. V).

### *A. How Runtime Uses the APIs*

When a runtime is created, the shared state of each NF along the configured service chain is initialized by calling $allocate_shared_state()$ and stored by a storage actor. After a flow actor is created to process a new flow, it first calls $allocate_new_fs()$ to create a flow state for each NF and stores these flow states throughout its lifetime. The flow actor processes a packet along the service chain by sequentially calling $process_pkt(input_pkt, fs, ss)$ for each NF, passing in the per-flow state, and shared state obtained from the storage actor. The shared state is sent back to the storage actor when the flow actor finishes processing the packet. When the flow actor expires (this is triggered by a per-actor timer), the flow actor first calls $flow_expires(fs, ss)$ for each NF to update the shared state and then executes some clean-up procedures, including calling $deallocate_fs(fs)$ to free the flow state. When a flow is migrated or recovered, the flow actor calls the last three APIs shown in Tbl. II to synchronize the flow state with the shared state for each NF, followed by executing some clean-up procedures.

### B. *Example NFs*

Using these APIs, we create four example NFs, i.e., a firewall, an intrusion prevention system (IPS), a load balancer and a NAT.

The firewall updates the connection information (per-flow state) and compare the 5-tuple of the flow with the access control list (shared state) to decide whether to drop the flow packet. The IPS scans the packet payload using an automaton (shared state) built with the Aho-Corasick algorithm [35], saves an index (per-flow state) to the current automaton state, and drops the flow packet if an attack signature is found. Since both shared states of the firewall and the IPS are read-only, i.e. the flow only reads the shared state without acquiring any resource from it, there is no need to implement the last three APIs in Tbl. II to synchronize the shared state.

The load balancer forwards each input flow to a server with the smallest workload among a set of backend servers. To achieve this, after selecting a server (per-flow state), the load balancer increases the workload counter (shared state) of the selected server to reflect the load balancing decision. Therefore, when the flow expires, or when the flow is migrated or recovered, the workload counter on that server should be properly decreased by implementing the last three APIs in Tbl. II.

The NAT operates by substituting the source IP address and source port of the flow packet with an allocated address (per-flow state) from a shared address pool (shared state). Within a cluster, the address pool of each NAT contains non-overlapping addresses. There is no need to implement the last 3 APIs in Tbl. II: we treat the address allocation from the address pool as persistent allocation that lasts throughout the lifetime of the flow, i.e., the flow only releases the address back to the address pool when it expires.

## V. System Management Operations

### A. *Fault Tolerance*

*1) Replicating Runtimes:* To perform lightweight runtime replication, we leverage the actor abstraction and state separation. In a runtime, important states associated with a flow are stored by the flow actor. The runtime can replicate each flow actor independently without check-pointing the entire container image [8], [9]. While achieving good throughput and fast flow recovery, this replication strategy also improves the packet processing delay and has good scalability, as each flow actor can replicate itself on another runtime without the need of dedicated back-up servers.

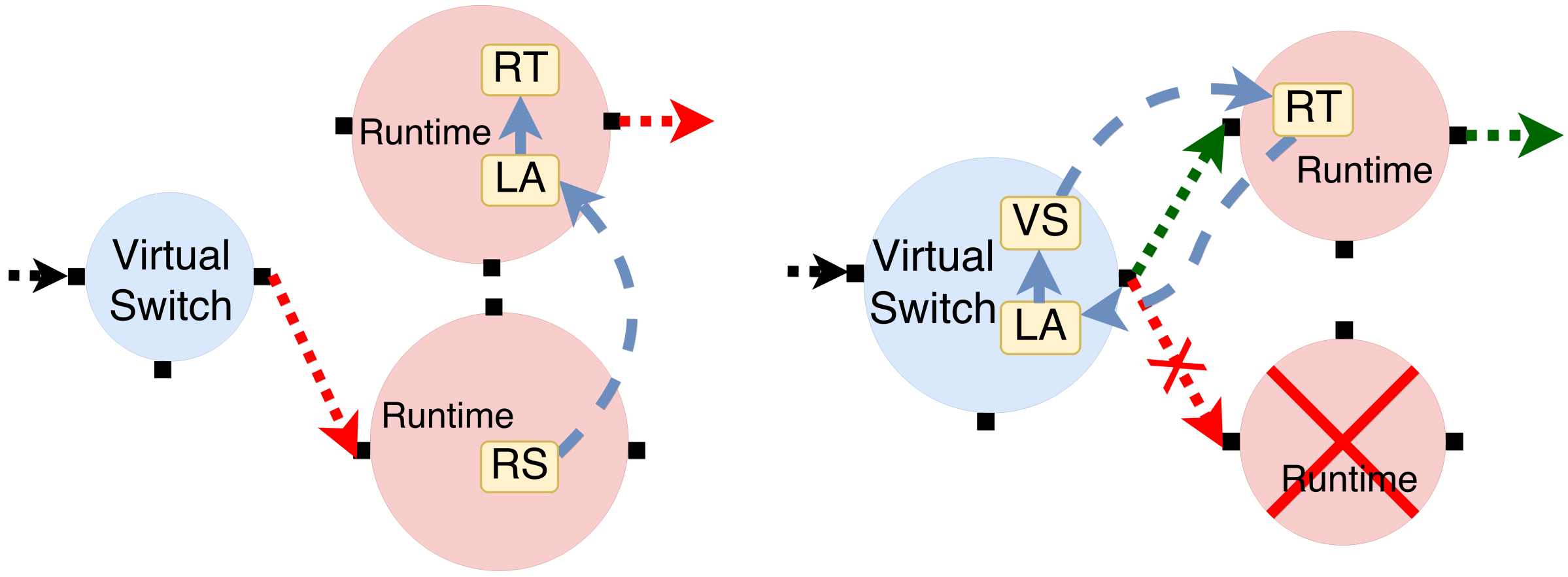

**(a)** Flow replication.

**(b)** Flow recover when the original runtime has failed.

**Fig. 3:** Flow replication and recovery: **RT** - replication target actor, **RS** - Replication source actor, **LA** - liaison actor, **VS** - virtual switch actor; **dotted line** - flow packets, **dashed line** - actor messages.)

The detailed flow replication process is illustrated in Fig. 3. When a runtime is launched, the coordinator sends a list of runtimes in the same cluster to its liaison actor via RPC $set_replicas(runtime_id_list)$. When a flow actor is created on the runtime, it acquires its replication target runtime from the liaison actor, selected in a round-robin fashion among all available runtimes received from the coordinator.

When a flow actor has finished processing a flow packet, it sends a replication actor message, containing the current flow states of all the NFs and the packet, directly to the liaison actor on the replication target runtime. The liaison actor further forwards the replication message to a replica flow actor sharing the same 5-tuple as the flow actor. The replica flow actor stores latest flow states contained in the message, and then directly sends the packet out, as shown in Fig. 3a. This design guarantees the same *output-commit* property as in [8]: the packet is sent out from the system only when all the state changes caused by the packet have been replicated.

The coordinator monitors the liveness of a runtime by sending heartbeat messages to the liaison actor of the runtime. When a runtime fails, the coordinator sends recovery RPC requests $recover(runtime_id)$ to all the runtimes containing replica flows of the failed runtime. When a runtime $R$ receives this RPC, it instructs each replica flow actor on runtime $R$ to send a request to the virtual switch actor, asking it to change the destination runtime to runtime $R$. After the acknowledge message from the virtual switch actor is received, the replica flow actor synchronizes the shared states by calling $flow_recover$ (Table II) and the flow is successfully restored on runtime $R$ (Fig. 3b).

*2) Replicating Virtual Switches:* Since a virtual switch is in fact a special runtime (Sec. III-C), the virtual switch can be replicated in the same way as described in Sec. V-A1. The only difference is that when the source virtual switch fails, the replica flow actors in the replication target virtual switch immediately become the primary flow actors without sending out a request to change the forwarding route. Instead, the coordinator takes control and updates the SDN rules to forward the input flows to the replication target virtual switch.

*3) Replicating Coordinator:* Since the coordinator is a single-threaded module, we can log and replicate information it maintains into a reliable storage system such as ZooKeeper[36]. The liveness of the coordinator is monitored by a guard process and it is restarted immediately in case of failure. On a reboot, the coordinator can reconstruct the system view by replaying logs.

### *B. Flow Migration*

Based on the actor model, flow migration in *NFVactor* can be regarded as a transaction between a source flow actor and a target flow actor, where the source actor delivers its entire state and processing tasks to the target actor. Flow migration is successful once the target actor has completely taken over packet processing of the flow. In case of unsuccessful flow migration, the source flow actor can fall back to regular packet processing and instruct to destroy the target actor.

In *NFVactor*, the coordinator starts flow migration by calling *set_migration_target* RPC method on a runtime, asking it to migrate a number of flows to another runtime. After receiving the ID of a migration target runtime, the flow actor starts migration by itself. The flow migration protocol used by flow actors is shown in Fig. 4, consisting of three request-response steps. In case of request timeout, the migration source actor performs clean-up procedures and reverts to normal packet processing.

**1st req-rep step:** The source flow actor sends 5-tuple of its flow to the liaison actor on the migration target runtime. The liaison actor creates a migration target actor using the 5-tuple, and sends a response back to the migration source actor. Meanwhile, migration source actor continues to process packets as usual.

**2nd req-rep step:** The source flow actor sends the 5-tuple of its flow and the ID of the migration target runtime to the liaison actor on the virtual switch. The liaison actor uses the 5-tuple to identify the virtual switch actor in charge and notifies it to change the destination runtime to the migration target runtime. After this change, the virtual switch actor sends a response back to the source actor, and the migration target actor starts to receive packets. Instead of processing the packets, the target actor buffers all the received packets until it receives the request in the 3rd step from the source actor. The migration source actor keeps processing received flow packets until it receives the response from the virtual switch.

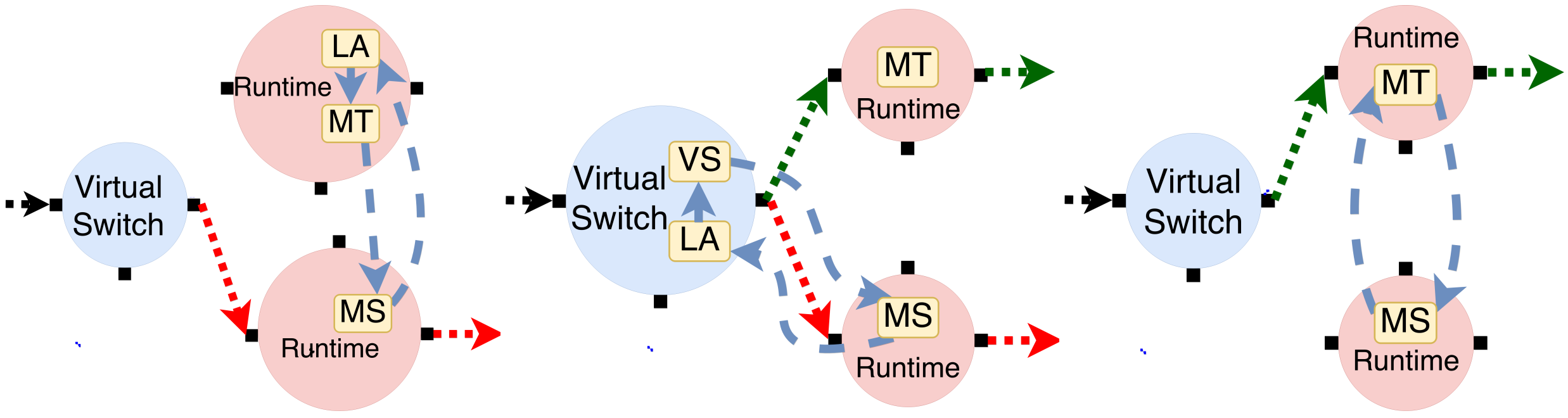

**Fig. 4:** The 3 flow migration steps: **MT** - migration target flow actor, **MS** - migration source flow actor, **LA** - liaison actor, **VS** - virtual switch actor; **dotted line** - flow packets, **dashed line** - actor messages.

**3rd req-rep step:** The source flow actor sends its flow states to the migration target actor. After receiving the flow states, the migration target actor saves them, calls $flow_migrate_in$ (Table II) to synchronize the shared states, and immediately starts processing all the buffered packets while sending a response to the source actor. The migration source actor calls $flow_migrate_out$ (Table II) to synchronize the shared states and then destroys itself.

*Loss Avoidance.* It is possible for our flow migration protocol to drop flow packets. However, packet drop caused by the flow migration protocol rarely happens in practice, even when concurrently migrating hundreds of thousands of flows (Sec. VII-D). We refer to this as loss-avoidance property.

In the 3rd step, before the request arrives at the migration target actor, the migration target actor has already received and buffered several flow packets. The buffer may overflow, causing migration target actor to drop several packets. However, our distributed flow migration finishes fast within several microseconds and such drop rarely happens in practice.

Still in the 3rd step, after the request is sent out by the source actor, it is possible for some flow packets to continue arriving at the source actor. These packets are sent out by the virtual switch

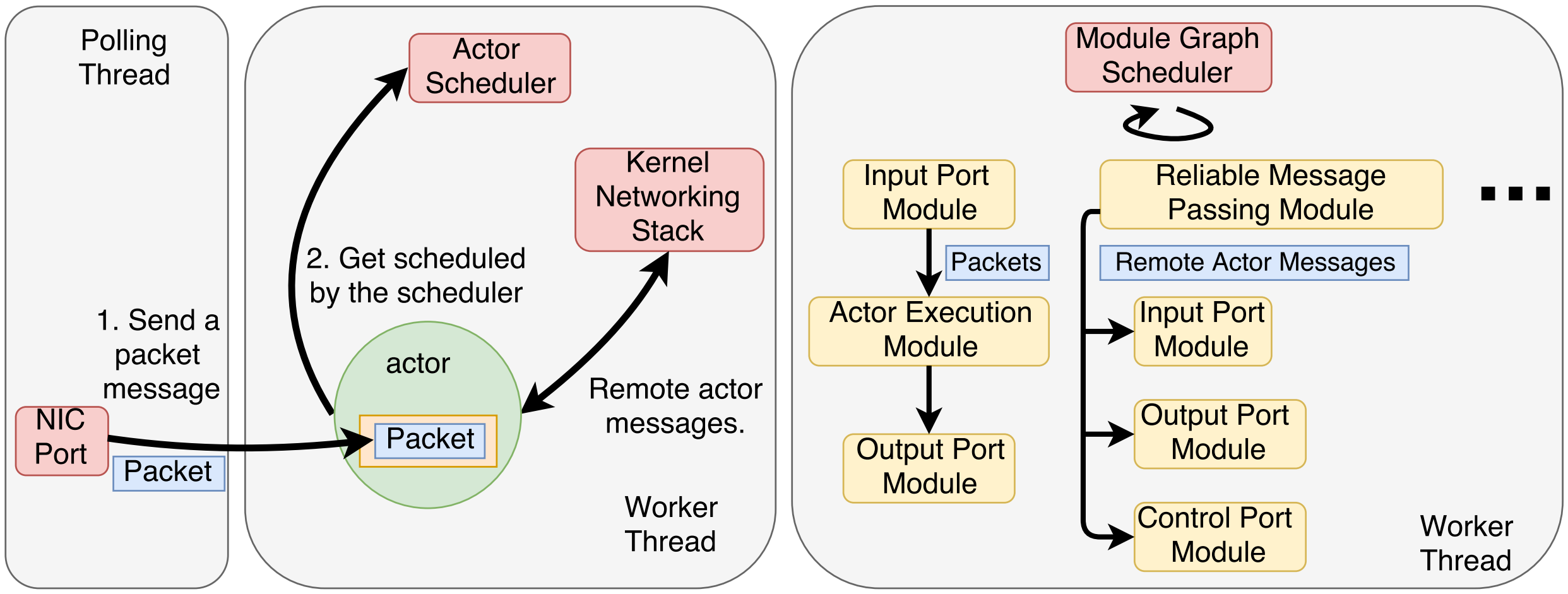

**(a)** Libcaf runtime. **(b)** *NFVactor* runtime.

**Fig. 5:** Comparison of runtime architecture.

actor before its destination runtime is changed in the 2nd step, but arrive later at the source actor than the response of the 2nd step. The source actor has to discard these packets because it has already sent out its flow state. *NFVactor* minimizes the occurrence of this problem by having the virtual switch actor transmitting the response message of the 2nd step in the same network path as the flow packets, so that we rarely observe this problem in practice.

It has been a common understanding that providing good properties for flow migration would trade off the performance of flow migration [15]. *NFVactor* mitigates this trade-off using distributed, high-performance flow migration based on the actor model.

## VI. Implementation

Throughout the development process of *NFVactor*, we initially chose libcaf [29] as the runtime, whose architecture is shown in Fig. 5a. However, the performance of libcaf [29] runtime is less than satisfactory, for both packet processing and resilience operations. We believe that the performance problems of libcaf runtime come from two aspects. *First*, the actor scheduler of libcaf is not suitable for handling IO of raw network packets. According to Fig. 5a, the scheduler schedules an actor run according to whether the actor has received a message. To inject dataplane packets into libcaf runtime, we have to set up a dedicated polling thread to poll the NIC port using DPDK [24] and send received packets to actors running in another worker thread by enqueueing the packet into actor's message queue. As verified in Sec. VII-A1, there is an expensive synchronization overhead between the polling thread and the worker thread, which decreases packet processing throughput by over 100%. *Second*, libcaf runtime still relies on inefficient kernel networking thread to exchange remote actor messages with other runtimes.

Realizing these problems, we design a customized actor runtime for *NFVactor* as shown in Fig. 5b, by leveraging two optimizations. *First*, we implement a module graph scheduler to schedule actors according to several pre-defined module graphs. The module graph scheduler combines various IO operations (polling NIC port, exchanging remote actor messages) with actor scheduling inside a single worker thread, effectively eliminating the thread synchronization overhead as in libcaf. *Second*, we bypass inefficient kernel networking stack and implement a high-performance, reliable message passing module running in user-space.

The tradeoff point when designing the customized runtime is programmability, as the programming interface exposed by the customized runtime is not as easy to use as the libcaf runtime. However, we believe that such a tradeoff is worthwhile due to improved performance.

### *A. Module Graph Scheduler*

Inspired by the scheduler design of BESS [25] and Click [37], the module graph scheduler keeps scheduling several module graphs to run in a round-robin fashion. A module graph consists of several processing modules, connected together into an acyclic graph. Inside a module graph, the actor messages are generated by a source module, flow through the connected module for processing before reaching the sink module, which consumes each message by either freeing it or sending it to the outside. Inside a module, the message handler of the corresponding actor is called for each actor message. The actor then pushes the processed message to the next connected module. When all the generated actor messages are consumed, the scheduler moves on to run the next module graph.

Fig. 5b illustrates two module graphs. The first module graph polls dataplane packets from the input port and generates packet messages, which are pushed along the module graph, processed by the flow actors and sent out from the output port. The second module graph fetches remote actor messages from the reliable message passing module and sends the remote actor message out from one of the three ports. There are two other module graphs that are used for receiving reliable actor messages and interacting with the RPC requests.

### B. Reliable Message Passing

Based on the module graph scheduler, we build a reliable message passing module, which inserts remote actor messages into a reliable packet stream for transmission. The module creates one ring buffer for each remote runtime and virtual switch. When a flow actor on this runtime sends a remote actor message, the module creates a packet, copies the content of the message into the packet and then enqueues the packet into the respective ring buffer. These packets are configured with a sequential number each, and appended with a special header to differentiate them from dataplane packets. When the second module graph in Fig. 5b is scheduled to run, the worker thread dequeues these packets from the ring buffers, and sends them to respective remote runtimes. A remote runtime acknowledges receipt of such packets. Retransmission is fired in case that the acknowledgement for a packet is not received after a configurable timeout (*e.g.*, 10 times the RTT in our current implementation). Running entirely in user-space, the performance of the reliable message passing module is good enough to saturate a 10Gbps link (Sec. VII-A2).

Since our goal is to reliably transmit remote actor messages over an inter-connected L2 network, we do not use user-level TCP [38], which may impose additional overhead for re-constructing byte streams into messages. In addition, packet-based reliable message passing provides additional benefits during flow migration and replication. Because the response in 2nd request-response step of flow migration is sent as a packet using the same path as the dataplane packets, reliable actor message passing enables us to implement loss-avoidance migration with ease (Sec. V-B).

## VII. Evaluation

We evaluate *NFVactor* using 4 Dell R430 servers, each equipped with an Intel Xeon E5-2650 CPU running at 2.30GHz with 20 logical CPU cores, 48GB memory and 2 Intel X710 10Gb NICs. The servers are connected through a 10GB Dell switch. In each server, the worker thread of each runtime is pinned to a dedicated logical core, while the RPC threads of all the runtimes are collectively pinned to logical core 0 to minimize the performance impact on the worker thread. To generate test traffic, we use a customized traffic generation module (the FlowGen module) of BESS [25], which is capable of generating input traffic up to 10Gbps (at around 14Mpps) with 64-byte packets.

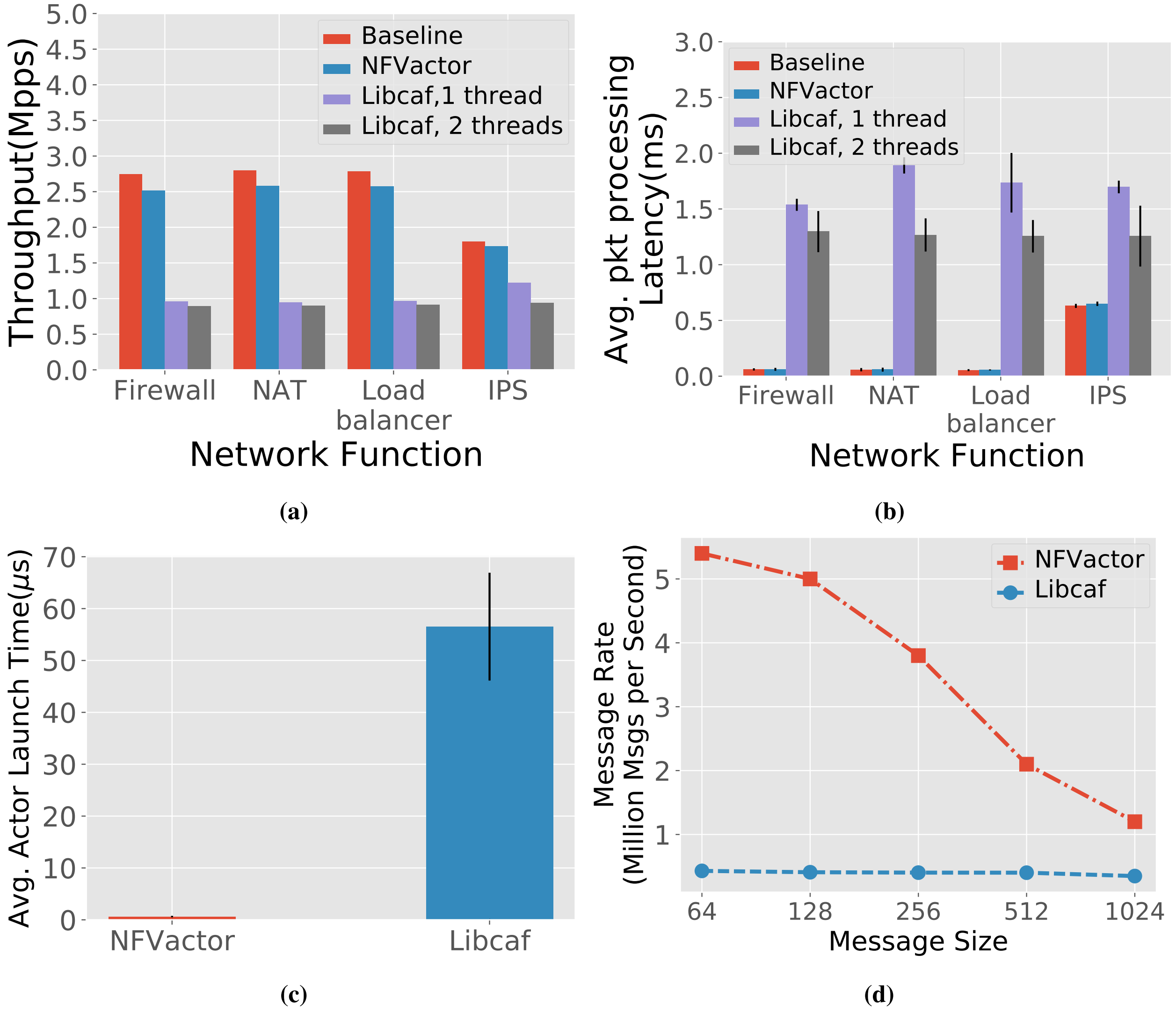

**Fig. 6:** Performance of the Runtime.

We use a single server to generate the traffic. We set up 6 virtual switches on another server, which are capable of handling input traffic at 10Gbps line rate and do not render bottlenecks. The rest of the servers are used to run runtimes.

### A. *Performance of the Runtime*

*1) Packet processing throughput and latency:* We first evaluate the packet processing throughput (number of packets processed per second) and latency (difference between the time that a packet enters the runtime to the time this packet is released from the runtime) of the *NFVactor* runtime by running the four implemented NFs. The traffic generator produces flows with 64-byte TCP packets, a 10pps (packets per second) flow rate and a 10-second active time. The overall packet rate of the input traffic is 5Mpps. For comparison, we also evaluate performance of libcaf runtime. We vary the number of worker threads used by the libcaf runtime to reflect the performance overhead of thread synchronization. Finally, we compare with four baseline NFs. The baseline NFs are implemented using a normal packet processing loop, sharing similar processing logic as the *process_pkt* API. The per-flow state in each baseline NF is stored in a fast hash table [39] without using the actor abstraction.

Fig. 6a and Fig. 6b show that *NFVactor* runtime achieves significantly larger throughput and smaller processing latency than libcaf runtime, and the performance of the NFs in *NFVactor* is close to that of the baseline NFs, as the actor abstraction does introduce a small overhead. According to Fig. 6a, when the number of the worker threads used by libcaf is increased, the total throughput drops by a small margin, due to increased synchronization overhead between the polling thread and multiple worker threads.

*2) Actor launch time and sending rate of remote actor messages:* The actor launch time is the time when the first packet of a flow is received to the time when this packet is processed by the launched actor. It is a strong performance indicator of the actor runtime system. In Fig. 6c, the input traffic is the same as in Sec. VII-A1. We see that the average actor launch time in *NFVactor* is much smaller than that of the libcaf runtime, as *NFVactor* pre-allocates flow actors into a ring buffer to speed-up actor launch time.

Fig. 6d shows the average sending rate of remote actor messages between two runtimes running on different servers. For various message sizes, the message rate achieved by *NFVactor* is significantly larger than that of libcaf. Especially, when the message size is larger than 256 bytes, we measure the consumed bandwidth of *NFVactor* to be around 9.1Gbps, which is close to the 10Gbps line rate. This result reflects that our user-space message passing module can significantly improve the performance of remote actor communication.

### *B. System Scalability*

We now evaluate the maximum packet processing throughput of *NFVactor* as the number of runtimes increases. We use two servers and configure the runtimes with service chain 'firewall (FW) → NAT → load balancer (LB)' (in one set of experiments) or service chain 'firewall (FW) → NAT → IDS' (in another set of experiments). To fully stress the system, we configure traffic generators to produce a mixture of short flows and long flows up to 10Gbps line rate. A short flow consists of 64-byte TCP packets with a 10pps packet rate and lasts for 1 second. A long

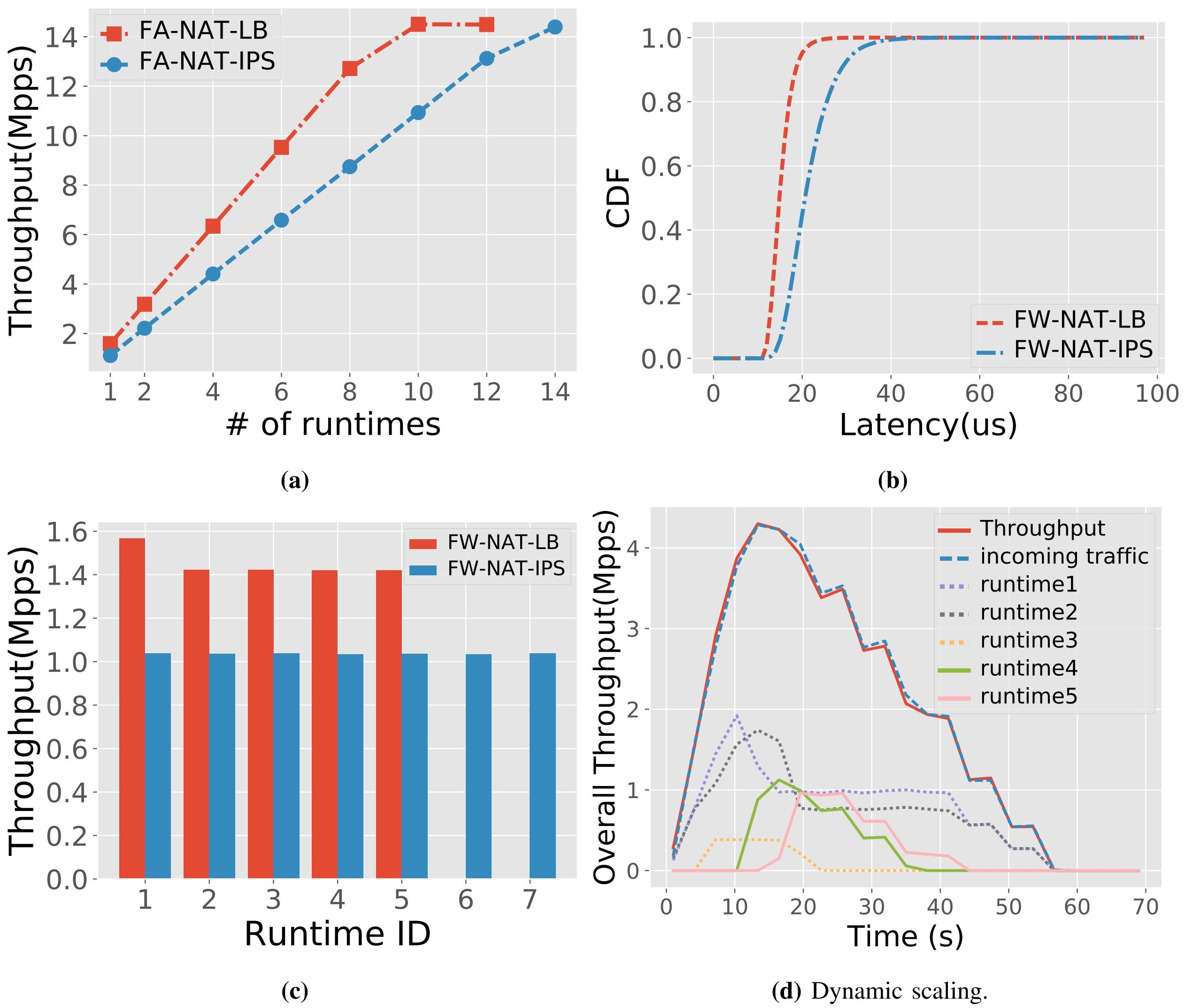

**(a)** **(b)** **(c)** **(d)** Dynamic scaling.

**Fig. 7:** System scalability.

flow consists of 64-byte TCP packets with a 10pps packet rate and lasts for 10 seconds. Each type of flow consumes half of generated bandwidth. We gradually increase the number of active runtimes and collect the total throughput achieved by the runtimes.

Fig. 7a shows the overall packet processing throughput increases linearly with the increase of the runtimes. Overall throughput of 14.49Mpps (9.70Gbps) and 14.39Mpps (9.67Gbps) are achieved when the runtimes run service chain ‘FW → NAT → LB’and ‘FW → NAT → IDS’, respectively. This verifies that *NFVactor* can approach 10Gbps line-rate packet processing for 64-byte small packets, even when the input traffic consists of many short-lived flows.

Fig. 7b shows the CDF of packet processing latencies, collected during a 20s period when 10 and 14 runtimes are used to run ‘FW → NAT → LB’ and ‘FW → NAT → IDS’, respectively.

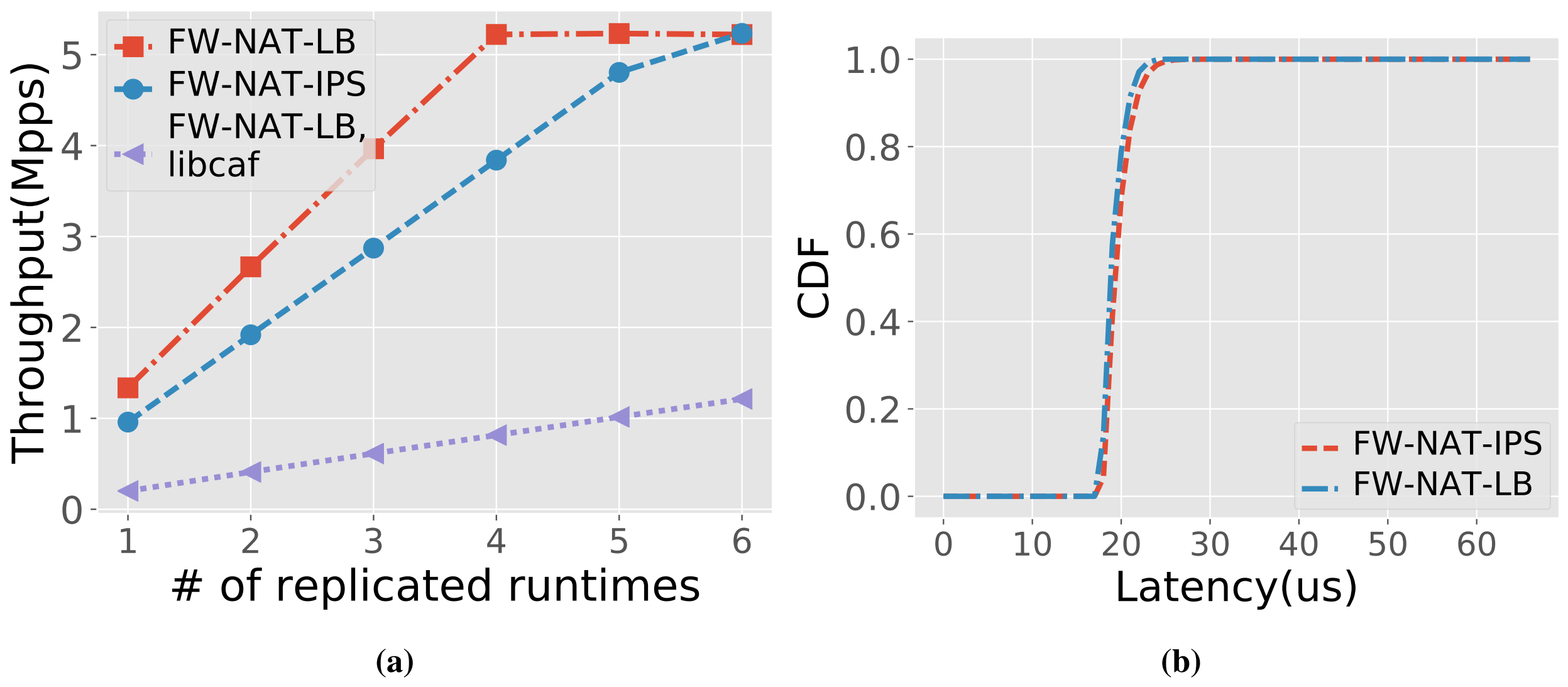

**Fig. 8:** Performance of flow replication.

The average latency for both service chain is around 20$\mu$s.

We next run the two service chains concurrently in the system. We run 5 runtimes in one server configured with 'FW $\rightarrow$ NAT $\rightarrow$ LB' and 7 runtimes in another server configured with 'FW $\rightarrow$ NAT $\rightarrow$ IDS'. The input traffic has a total packet rate of 14.50Mpps and shares the same mixed pattern to produce Fig. 7a. The input traffic is evenly split among the two service chains. Fig. 7c shows the throughput of each runtime. We can see that a total throughput of 7.25Mpps can be reached by each service chain. The workload is also evenly balanced among runtimes in the same server, demonstrating the effectiveness of our virtual switches for load balancing under mixed short and long flows.

Finally, the performance of the dynamic scaling controlled by the coordinator is shown in Fig. 7d. The traffic generator generates traffic of 40K flows with increasing packet rate for the first 15 seconds, and then gradually decreases the packet rate of the traffic for the last 45 seconds. The cluster is initialized with two runtimes (runtime 1 and 2) running 'FW $\rightarrow$ NAT $\rightarrow$ IDS' service chain. Starting from the 10th second, runtimes 1 and 2 become overloaded and their workload is gradually migrated away to runtime 4 and 5. With dynamic scaling, the achieved throughput can correctly match the input traffic during the 60 seconds.

## *C. Performance of Flow Replication*

In this set of experiments, input flows are produced following the same mixed pattern to produce Fig. 7a. The coordinator chooses two servers and launches the same number of runtimes on them. The coordinator instructs each runtime on a server to replicate its flows to a distinct runtime in another server. We gradually increase packet rate of the input traffic and the number of runtimes running on each server, to investigate throughput and scalability when flow replication is enabled.

**TABLE III:** Recovery time and # of flows recovered

| | FW→NAT→LB | FW→NAT→IDS | FW→NAT→LB, libcaf |
|---|---|---|---|
| Average recovery time for each runtime | 65.3ms | 66.6ms | 934.2ms |
| Number of flows recovered on each runtime | at least 87k flows | at least 87k flows | at least 20k flows |

Fig. 8a shows that both service chains can scale up to handle the handle the maximum replication packet rate of 5.22Mpps when six runtimes running on a server concurrently replicate their traffic to six replica runtimes on another server. We can see that the maximum replication packet rate can not reach the line rate, which is around 14.4Mpps for 64-byte packets. This is because when the replication throughput reaches 5.22Mpps, the bandwidth for transmitting replication messages reaches around 10Gbps, fully saturating the link for transmitting the replication messages. If the bandwidth of this link is increased to 40Gbps or more, the maximum replication throughput achieved by *NFVactor* can be further improved. Finally, when the libcaf version of implementation is used, the replication throughput is significantly lower.

Fig. 8b shows the CDF of packet processing latencies of *NFVactor* when flow replication is enabled. The latency measured in this experiment is difference between the time that the packet enters replication source runtime to the time this packet is released from the replication target runtime. For both service chains, the number of runtimes on each of the two servers is 6 while the input packet rate is 5.22Mpps. The average latency is around 20$\mu$s for both service chains.

Table III shows the average recovery time of 6 replication target runtimes when their corresponding replication source runtimes fail simultaneously. We can see that *NFVactor* has a much shorter recovery time than the libcaf version even when processing a larger number of flows.

*1) Comparison with FTMB:* We compare the performance of flow replication in *NFVactor* with the reported performance of FTMB [8]. Both systems achieve throughput up to millions of packets per second and recovery time of tens of milliseconds with flow replication enabled. *NFVactor* has a more stable packet processing latency (according to Fig. 8b, smaller than 70

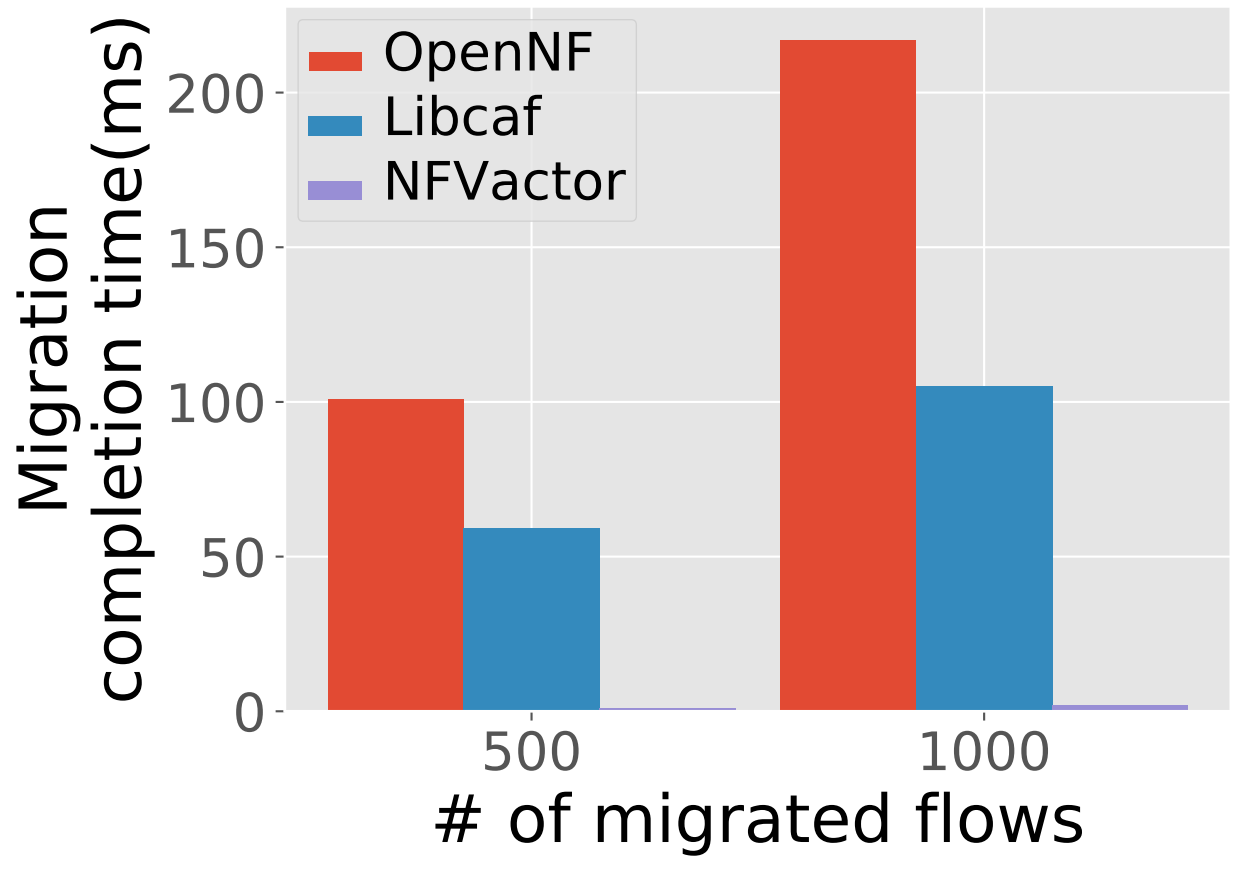

**(a)**

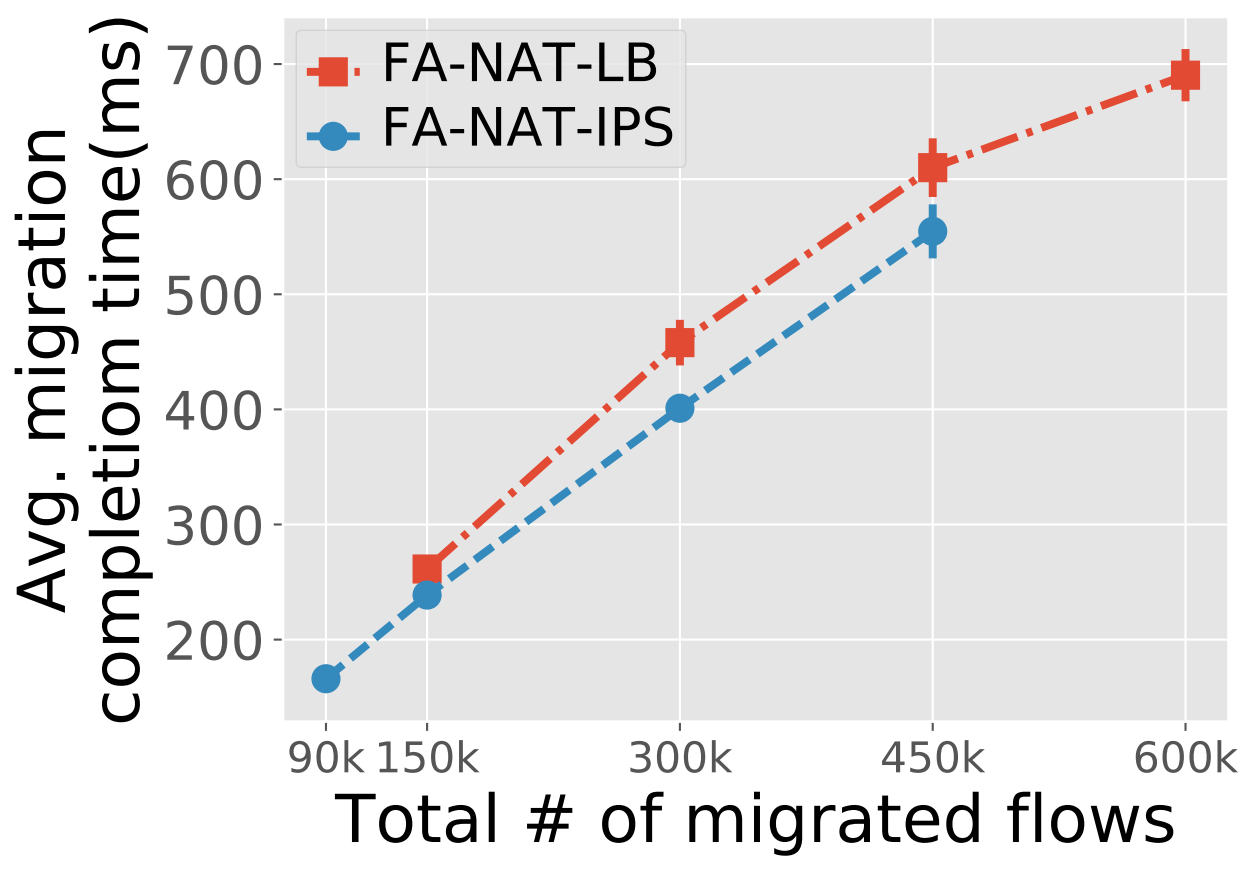

**(b)**

**Fig. 9:** Flow migration completion time.

microseconds, with an average of 20 microseconds) because it does not need to checkpoint the runtime, whereas FTMB introduces a relatively high packet processing latency (up to 3000 microseconds) when checkpointing kicks in.

### *D. Performance of Flow Migration*

We first compare flow migration performance among *NFVactor*, libcaf runtime, and OpenNF. Both *NFVactor* and libcaf runtime run the example firewall. We also port the example firewall to work with OpenNF. We send the same number of flows to the three firewalls and each flow has a 10pps packet rate. In Fig. 9a, the time to migrate the respective number of flows is much smaller with *NFVactor* (0.7ms for 500 flows and 1.5ms for 1000 flows), when compared to OpenNF (101ms for 500 flows and 217ms for 1000 flows) and the libcaf runtime (59ms for 500 flows and 105ms for 1000 flows). Due to efficient runtime design, *NFVactor* can out-perform OpenNF by 144 times when migrating 1000 flows.

We next show the time taken for concurrently migrating a large number of flows among multiple pairs of runtimes. The traffic generator produces a number of flows with a 10pps flow rate, each lasting for 1 minute. The flows are first sent to three runtimes running in one server. Then the coordinator instructs the three runtimes to concurrently migrate all the flows to three runtimes running in another server.

Fig. 9b shows the average migration completion time of the three migration source runtimes. The standard deviation of the completion time is shown as error bar in Fig. 9b as well. *NFVactor* can migrate 600K flows (with 6Mpps total throughput) from three migration source runtimes running in one server to three migration target runtimes running in another server using around 700ms. Besides the performance boost enabled by efficient runtime design, the decentralized flow migration also contributes to the performance. Since the flow migration are concurrently carried out among three pairs of runtimes, each pair of runtime only needs migrates around 200K flows. This significantly reduce the eventual flow migration completion time for all the 600K flows. If the 600K flows are sequentially migrated, the resulted flow migration completion time may be prolonged to over 2 seconds. Finally, throughout the evaluation in Fig. 9b, we observe zero packet loss caused by our flow migration protocol.

## VIII. Related Work

Since the introduction of NFV [1], a broad range of studies have been carried out, for bridging the gap between specialized hardware and network functions [40], [41], [42], [43], scaling and managing NFV systems [6], [2], flow migration [14], [44], [15], NF replication [9], [8], and traffic steering [45]. In these work, NF instances are typically created as software modules running on separate VMs or containers. *NFVactor* customizes a runtime system to run flow actors and embeds service chain processing within the flow actor abstraction, to achieve transparent and highly efficient failure resilience guarantee.

Among the existing studies, StatelessNF [34] shares similar design goals with *NFVactor*, *i.e.*, to enable transparent system scalability and failure resilience. However, the methodology used by StatelessNF is orthogonal to that of *NFVactor*: it stores flow states in a reliable database [46] to achieve failure resilience, while *NFVactor* exploits the actor model. Compared with StatelessNF, *NFVactor* can approach line-rate packet processing and does not rely on RDMA equipment.

Besides, the Click modular router [37] is the first work to introduce modular design for simplifying NF construction. The module graph scheduler used by *NFVactor* is partially inherited from the scheduler of Click. However, *NFVactor* uses such a scheduler to speed up actor processing. Flurries [47] proposes fine-grained per-flow NF processing, by dynamically assigning a flow to a lightweight NF. Sharing some similarities, *NFVactor* enables micro service chain processing of each flow in one actor, but focuses on providing transparent failure tolerance based on the actor model. OpenBox [3] merges the processing logic of multiple VNFs, therefore improving the modularity and processing performance. Even though *NFVactor* uses the traditional sequential service chain, we believe that the flexibility of actor model can help us adopt the idea of OpenBox in *NFVactor*, which we leave for future exploration. StateAlyzr [44] uses static analysis to automate flow state extraction and simply human effort for enabling flow migration. However, enabling high-performance flow migration still requires a holistic design like *NFVactor*.

## IX. Conclusions and Discussions

We have presented *NFVactor*, an NFV system using actor model to achieve transparent and highly efficient failure resilience. *NFVactor* advocates a novel one-flow-one-actor principle to improve the parallelism and performance of resilience operations, while the efficiency of the actor model is guaranteed by a high-performance runtime. Our experiments show that *NFVactor* achieves good scalability and high packet processing speed, as well as fast flow migration and failure recovery.

Powered by a completely new architecture, *NFVactor* does require NFs to be rewritten or ported using the provided APIs. We believe that there will be a growing need for implementing new NFs with further adoption of the NFV paradigm and the flexible API design of *NFVactor* makes it possible to port a wide range of existing NFs. Besides, our holistic design approach will be among the method that NF developers may choose from, valuable for decoupling complexity of critical system services from the core logic of NF.